\documentclass{emulateapj}
\usepackage{apjfonts}

\newcommand{\te}{t_{\rm E}}

\newcommand{\thetae}{\theta_{\rm E}}
\newcommand{\rvec}{\mbox{\bf r}}

\shortauthors{HAN ET AL.}
\shorttitle{BLENDING IN MICROLENSING}

\begin{document}

\title{Blending in Future Space-based Microlensing Surveys}

\author{
Cheongho Han\altaffilmark{1},
Byeong-Gon Park\altaffilmark{2}, 
Ho-Il Kim\altaffilmark{2}, and
Kyongae Chang\altaffilmark{3}
}

\altaffiltext{1}{Program of Brain Korea 21, Department of Physics, 
Institute for Basic Science Research, Chungbuk National University, 
Chongju, Korea 361-763; cheongho@astroph.chungbuk.ac.kr}
\altaffiltext{2}{Korea Astronomy and Space Science Institute, Hwaam-Dong, 
Yuseong-Gu, Daejeon, Korea 305-348; bgpark,hikim@kasi.re.kr}
\altaffiltext{3}{Department of Physics, Chongju University, Chongju, 
Korea 360-764; kchang@chongju.ac.kr}


\begin{abstract}
We investigate the effect of blending in future gravitational microlensing 
surveys by carrying out simulation of Galactic bulge microlensing events 
to be detected from a proposed space-based lensing survey.  From this 
simulation, we find that the contribution of the flux from background 
stars to the total blended flux will be equivalent to that from the lens 
itself despite the greatly improved resolution from space observations, 
implying that characterizing lenses from the analysis of the blended flux 
would not be easy.  As a method to isolate events for which most of the 
blended flux is attributable to the lens, we propose to use astrometric 
information of source star image centroid motion.  For the sample of 
events obtained by imposing a criterion that the centroid shift should be 
less than three times of the astrometric uncertainty among the events for 
which blending is noticed with blended light fractions $f_{\rm B}>0.2$, 
we estimate that the contamination of the blended flux by background stars 
will be less than 20\% for most ($\sim 90\%$) of the sample events.  The 
expected rate of these events is $\gtrsim 700$ events/yr, which is large 
enough for the statistical analysis of the lens populations.
\end{abstract}

\keywords{gravitational lensing }

\section{Introduction}

Observational experiments to detect light variations of stars induced 
by gravitational microlensing have been and are being carried out for 
more than a decade \citep{udalski93, alcock93, aubourg93, bond01}.  
Originally, the experiments were initiated for the purpose of searching 
for dark matter in the Galactic halo \citep{paczynski86}.  However, 
microlensing developed many applications in various aspects of galactic, 
stellar, and planetary astrophysics such as probing the structure of the 
inner galaxy \citep{kiraga94, evans94, han95b, zhao95}, resolving the 
photosphere of distant stars \citep{witt95, albrow99, abe03, field03}, 
and detecting extrasolar planets \citep{mao91, gould92, bond04, udalski05, 
beaulieu06, gould06}.  Current experiments are routinely detecting more 
than 500 events every season and the total number of detected events is 
now reaching 3000.

The probability of a star being gravitationally lensed is very low (of 
the order of $10^{-6}$).  As events are very rare, lensing searches are 
conducted toward very crowded star fields such as the Galactic bulge in 
order to monitor as many stars as possible within a single frame.  For 
observations in such a dense field, blending of stellar images is inevitable 
\citep{distefano95, wozniak97}.  The lightcurve of a blending-affected 
event is expressed as 
\begin{equation}
F_{\rm obs} = AF_0 + F_{\rm B};\qquad A={u^2+2\over u(u^2+4)},
\label{eq1}
\end{equation}
where $F_0$ is the baseline flux of the lensed source star, $F_{\rm B}$ 
is the amount of the blended flux, $A$ is the magnification induced by 
lensing, and $u$ is the lens-source separation normalized by the Einstein 
ring radius $\thetae$.  The normalized lens-source separation is related 
to the lensing parameters by
\begin{equation}
u = \left[ \left( {t-t_0\over \te}\right)^2-u_0^2 \right]^{1/2},
\label{eq2}
\end{equation}
where $\te$ is the Einstein timescale, $t_0$ is the time of the closest 
lens-source approach, and $u_0$ is the separation at that moment (impact 
parameter).  Among the lensing parameters, the Einstein timescale is 
related to the mass and location of the lens and relative lens-source 
transverse speed, and thus it can provide constraint about the lens 
system.  However, the constraint is weak because the timescale results 
from the combination of the physical parameters.  Blending aggravates 
the photometric precision due to the increased noise from the blended 
flux.  In addition, blending causes accurate determinations of the 
lensing parameters difficult due to the poorly known baseline flux of 
the source star.  As a result, the constraint about 
the lens system becomes further weaker.

Recently, next-generation microlensing experiments based in space are 
being seriously discussed.  {\it Microlensing Planet Finder} ({\it MPF}), 
that succeeds the original concept of {\it Galactic Exoplanet Survey 
Telescope} ({\it GEST}) mission \citep{bennett02}, is a space mission 
proposed to NASA's Discovery Program  with the main goal of searching 
for a large sample of extrasolar planets by using gravitation lensing 
method \citep{bennett04}.  By conducting observations in space, it is 
possible to minimize blending effect thanks to the improved resolution.  
Another important advantage of the space observation may be that one can 
directly detect and characterize the lens based on the observed flux.  
This is possible when a light-emitting star is responsible for a lensing 
event and the flux from the lens contributes a substantial fraction of 
the observed flux.  It is believed that majority of events detected 
toward the Galactic bulge are caused by stars \citep{han03}.  In 
addition, the {\it MPF} mission plans to monitor faint main-sequence 
stars in order to optimize detections of low-mass planets by minimizing 
finite-source effect.  Therefore, the light from the lens would not be 
negligible for a significant fraction of the event sample, and thus the 
existence of the lens could be revealed by the excess flux from the lens.  
By conducting multi-band observations, it would be possible to determine 
the spectral type of the lens, which would provide much tighter 
constraint about the lens.

However, this new method of constraining lenses can be applicable under 
the condition that source stars to be observed are in most cases free 
from blending by the flux of blends other than lenses.  If an unrelated 
background star is located within the resolution disk of the lensed 
source star, the blended flux comes not only from the lens but also 
from the background star.  This causes complexity in the analysis of 
the observed blended light, making it difficult to uniquely characterize 
the lens.  \citet{han05} (hereafter Paper I) warned of the possible 
contamination of the blended flux by background stars.  As a method to 
resolve the contamination, he suggested to use the astrometric 
information of the source star image centroid shift.  If an event is 
affected by background stellar blending, the centroid will shift from 
the apparent position of the combined image of the source plus blended 
stars toward the lensed star during the event \citep{goldberg98}.

In this paper, we further investigate the blending problem in future 
space-based lensing surveys following Paper I and probe the feasibility 
of sorting out events whose blended flux is mostly attributable to the 
lenses by using the additional astrometric information of the centroid 
shift.  For this, we carry out simulation of Galactic bulge microlensing 
events expected to be detected from the proposed {\it MPF} survey by 
imposing realistic observational conditions and detection criteria.  
Based on this simulation, we investigate the expected pattern of blending 
and the seriousness of the contamination by background stars.  We also 
estimate the fraction of events that can be identified to be free from 
background stellar blending by using the proposed method.

\section{Simulation}

The basics scheme of the simulation of lensing events is similar to the 
one in Paper I.  The locations of the lenses and source stars are allocated 
based on \citet{han95a} Galactic mass distribution model, that is composed 
of a double-exponential disk and a barred bulge.  The velocity distribution 
of the bulge is deduced from the tensor virial theorem, while the disk 
velocity distribution is modeled to have a flat rotation speed of 
$v_c=220\ {\rm km}\ {\rm s}^{-1}$ and velocity dispersion along and normal 
to the disk plane of $\sigma_\parallel=30\ {\rm km}\ {\rm s}^{-1}$ and
$\sigma_\perp=20\ {\rm km}\ {\rm s}^{-1}$, respectively.  The brightnesses 
of the bulge stars are assigned on the basis of a combined luminosity function 
of constructed based on those of \citet{holtzman98} and \citet{gould97} 
considering extinction and distance modulus.  The lens masses are assigned 
based on the mass function of \citet{gould00}, which is composed of stars 
and stellar remnants.  The mass function model is constructed under the 
assumption that bulge stars are formed initially according to a double 
power-law distribution of $dN/dM=k (M/0.7\ M_\odot)^\gamma$, where 
$\gamma=-2.0$ for $M\geq 0.7\ M_\odot$ and $\gamma=-1.3$ for $M<0.7\ M_\odot$.
Based on this initial mass function, remnants are modeled by assuming that 
the stars with initial masses $1\ M_\odot < M < 8\ M_\odot$, $8\ M_\odot 
< M < 40\ M_\odot$, and $M>40\ M_\odot$ have evolved into white dwarfs 
(with a mean mass $\langle M\rangle=0.6\ M_\odot$), neutron stars ($\langle 
M\rangle=1.35\ M_\odot$), and black holes ($\langle M\rangle=5\ M_\odot$), 
respectively.  For events caused by stellar lenses, the lens brightness is 
determined based on the lens mass by using a mass-luminosity relation.  
For events caused by remnant lenses, the lens is assumed to be dark.

However, the main focus of Paper I was investigating possible contamination 
of blended flux in general and thus it is not focused on a specific 
survey.  As a result, there are several major differences between the 
simulations in Paper I and this work.  The differences are described below.
\begin{enumerate}
\item
Since no specific instrument was defined, the analysis in Paper I was 
based on all possible events regardless of their detectability.  In this 
work, analysis are based only on events that could be detected from the 
{\it MPF} survey.  The detectability of an events is subject to the 
condition and strategy of observation.  For the observation field, we 
assume that the future survey will monitor toward a field centered at 
$(l,b)\sim (1.2^\circ, -2.4^\circ)$ with a field of view of $(\Delta l, 
\Delta b) \sim(0.93^\circ, 2.8^\circ)$.  For a sampling frequency, we 
assume that the survey will continuously observe the target field with 
an interval of 15 minutes during 9 months of a year.  For each interval, 
we assume that a 10-minute exposure image is acquired with a photon count 
rate of 13 photons/sec for an $I=22$ star.
\item
Another factor that determines the event detectability is the selection 
criteria.  We judge the detectability of an event based on the uncertainties 
of the lensing parameters recovered from the lightcurve.  We determine 
the uncertainties by computing the curvature matrix of $\chi^2$ surface 
of the lensing parameters for each event produced by the simulation.  
For the case of a lensing lightcurve, the curvature matrix is defined as 
\begin{equation}
b_{ij}=\sum_{k}^{N_{\rm obs}}
{\partial F_{{\rm obs},k}\over \partial p_i}{\partial F_{{\rm obs},k}
\over \partial p_j} {1\over \sigma_k^2},
\label{eq3}
\end{equation}
where $N_{\rm obs}$ is the number of observations, $\sigma_k=
\sqrt{F_{\rm obs}}$ is the photometric precision of each measurement, 
and $p_i\equiv (F_0, F_{\rm B}, u_0, t_0, \te)$ are the five lensing 
parameters required to fit the lightcurve of a standard point-source 
single-lens event.  Then, the uncertainties of the individual lensing 
parameters correspond to the diagonal components of the inverse curvature 
matrix (covariance matrix), i.e.
\begin{equation}
\sigma_{p_i} = \sqrt{c_{ii}};\qquad c\equiv b^{-1}.
\label{eq4}
\end{equation}
With the determined uncertainties, we set the detection criteria of events 
such that the fractional uncertainties of the timescale and blended flux 
should be less than 20\%, i.e.\  $\sigma_{t_{\rm E}}/ \te \leq 0.2$ and 
$\sigma_{F_{\rm B}}/F_{\rm obs}\leq 0.2$.  Under this definition of event 
detectability, we note that events can be detected even when the source 
star trajectory does not enter the Einstein ring of the lens.  This can 
be seen in the upper left panel of Figure~\ref{fig:one}, where we present 
the distribution of impact parameters of events.  From the distribution, 
one finds that a fraction of events have impact parameters larger than 
unity, especially for events involved with relatively bright source stars 
($I_0\lesssim 22$).  In Paper I, detectable events were limited only to 
those with source trajectories entering the Einstein ring, i.e.\ $u_0<1.0$.

\begin{figure}[t]
\epsscale{1.2}
\plotone{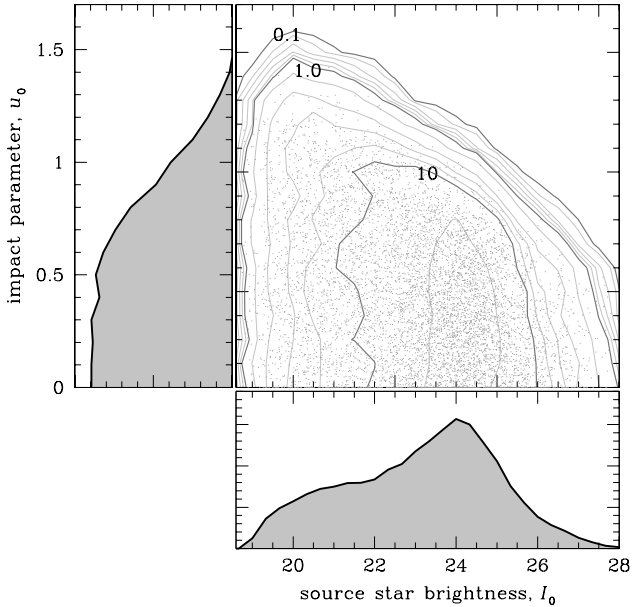}
\caption{\label{fig:one}
Distribution of lensing events in the parameter space of the $I$-band
baseline source brightness, $I_0$, and lens-source impact parameter, 
$u_0$.  The distribution is for events to be detectable from a lensing 
survey equipped with a space telescope similar to the {\it Microlensing 
Space Finder} mission.  The contour levels are in relative scale.
}\end{figure}

\item
Another major difference is the brightness range of source stars.  In 
Paper I, we considered only stars with $I$-band baseline brightness of
$I_0\lesssim 23$.  With the implementation of the detection criteria 
described above, we found that events associated with much fainter 
stars could be detected.  In this work, we, therefore, do not set the 
limit of source star brightness as long as the event meets the detection 
criteria.  With the new criterion, we find that events can be detected 
with source brightness down to $I_0\sim 27$, as shown in the lower panel 
of Figure~\ref{fig:one}, where the distribution of $I$-band baseline 
source star brightness is presented.  With this new range of the source 
star brightness, the baseline flux of the source star decreases in average 
while the blended flux remains similar to the previous estimation.  As 
a result, the average value of the blended light fraction, $f_{\rm B}= 
F_{\rm B}/ (F_0+F_{\rm B})$, is smaller than the value in Paper I.

\item
The last major change is the definition of the resolution angle 
$\theta_{\rm res}$.  In Paper I, we assume that two stars cannot be 
resolved if the separation between them is less than a fixed value of 
the diffraction limit (DL) regardless of the flux ratio between the 
two stars.  To describe the experiment more realistically by considering 
the variation of the resolution angle depending on the relative flux, we 
set the resolution angle as
\begin{equation}
\theta_{\rm res} = 
\cases
{
0.5\ {\rm FWHM}     & if $F_2/F_1\leq 0.1$,  \cr
[-0.44 (F_2/F_1)+0.54]\ {\rm FWHM}   & otherwise,  \cr
}
\label{eq5}
\end{equation} 
where $F_1$ and $F_2$ are the fluxes of the brighter and fainter stars,
respectively, and ${\rm FWHM}\sim 2\times {\rm DL}\sim 0.4''$ corresponds 
to the diameter of the resolution disk.  Then, the resolution angle for 
two equally bright stars is $\theta_{\rm res}= 0.1\ {\rm FWHM}\sim 
0.04''$.\footnote{For the {\it MPF} mission, the pixel size of the 
detector is $\sim 0.24''/{\rm pixel}$, which is bigger than the assumed 
${\rm DL}=0.2''$.  However, we note that images to be taken by the 
{\it MPF} will be dithered and there will be many of them.  Therefore, 
it would be possible to create drizzled images that recover the 
diffraction limit.}  With the adoption of the new definition of the 
resolution angle, the average blended light fraction further decreases 
compared to that in Paper I.
\end{enumerate}

\begin{deluxetable}{lr}
\tablecaption{{\it MPF} Event Rate \label{table:one}}
\tablewidth{0pt}
\tablehead{
\colhead{categories} &
\colhead{rate} }
\startdata
total event rate  & 3415/yr \ \ \ \ \ \ \ \ \ \ \ \ \ \   \\
\ \ \ \  -- bulge self-lensing events  & 2192/yr (64.2\%)  \\
\ \ \ \  -- disk-bulge events          & 1223/yr (35.8\%)  \\
\ \ \ \  -- stellar lens events        & 2510/yr (73.5\%)  \\
\ \ \ \  -- remnant events             & 905/yr  (26.5\%)  \\
\enddata
\tablecomments{ 
Detection rate of microlensing events from the {\it MPF} survey.
}
\end{deluxetable}

Based on the simulation, we estimate that the total event rate of the 
{\it MPF} survey would be $\Gamma_{\rm tot}\sim 3500\ {\rm events/yr}$, 
among which bulge self-lensing and disk-bulge events comprise $\sim 64\%$ 
and $\sim 36\%$, respectively.  The ratio between the events caused by 
stellar and remnant lenses is $\Gamma_\star : \Gamma_{\rm rem}=73.5: 26.5$.
We summarize the result of event rate in Table~\ref{table:one}.

\section{Results}

\begin{figure}[t]
\epsscale{1.2}
\plotone{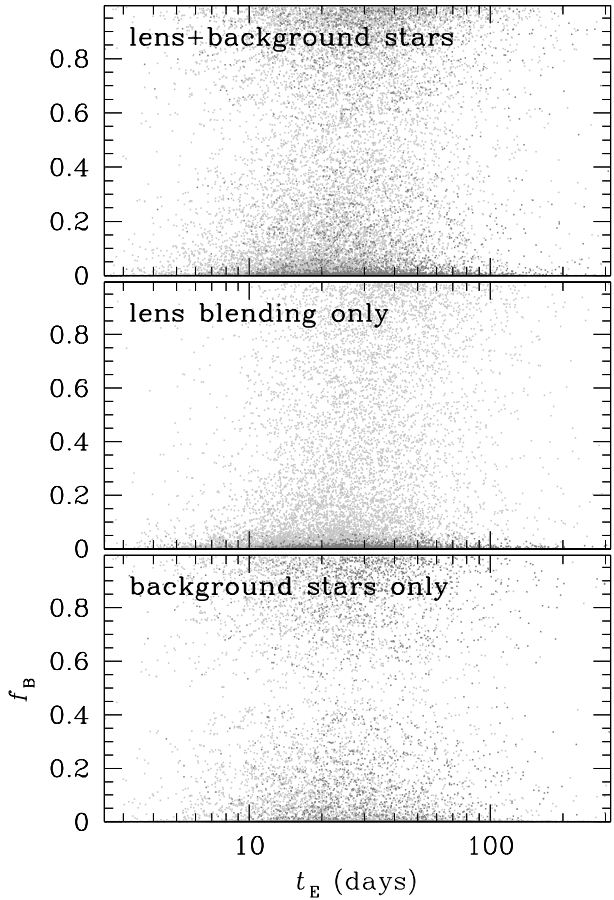}
\caption{\label{fig:two}
Distribution of microlensing events detectable from the {\it MPF} survey 
in the parameter space of the Einstein timescale, $t_{\rm E}$, and the 
blended light fraction, $f_{\rm B}$ (top panel).  The other two panels 
are the distributions expected when the only blending source is the lens 
(middle panel) and background stars (bottom panel), respectively.  In each 
panel, the points marked by bright and dark-tone greyscales represent the 
events caused by stellar and remnant lenses, 
respectively.
}\end{figure}

The top panel of Figure~\ref{fig:two} shows the distribution of events 
expected to be detected from the {\it MPF} survey in the parameter space 
of the Einstein timescale and blended light fraction.  We compare the 
distribution with those of events expected when the only blending source 
is the lens (middle panel) and background stars (bottom panel).  In each 
panel, the points marked by bright and dark-tone greyscales represent 
events caused by stellar and remnant lenses, respectively.

In the case where the dominant source of blending is the lens itself, 
the distribution has the following distinctive features.  
\begin{enumerate}
\item
First, the region with very little blended light fraction ($f_{\rm B}
\lesssim 0.05$) is densely populated by events caused by remnant lenses, 
which are dark and thus have no contribution to the blended flux.  
\item
Second, except for the ones caused by remnants, events are distributed 
smoothly over the entire range of the blended light fraction.  
\item
Third, there exists a correlation between the timescale and blending 
fraction.  This correlation arises because heavier lenses tend to be 
brighter and thus events with longer timescales are more likely to be 
affected by larger amount of blended flux.  We find that the mean 
timescale is $\langle \te \rangle\sim 20$ days for stellar-lens events 
with $f_{\rm B} \leq 0.1$, while $\langle \te \rangle \sim 35$ days for 
events with $f_{\rm B}\geq 0.9$.  However, this correlation is not very 
strong due to the large dispersions of the Einstein timescale and lens 
brightness.
\end{enumerate}

\begin{figure}[t]
\epsscale{1.2}
\plotone{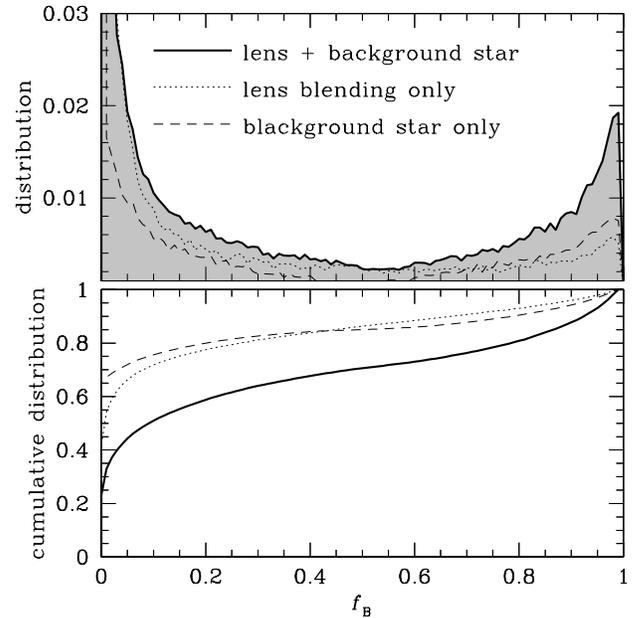}
\caption{\label{fig:three}
Distribution of blended light fractions, $f_{\rm B}$, for microlensing 
events to be detectable from the {\it MPF} survey (solid curve).  The 
dotted and dashed curves are the distributions when the only blending 
source is assumed to be the lens and background stars, respectively.  
Lower panel shows the cumulative distributions.
}\end{figure}

In the case where most of the blended flux comes from background stars, 
on the other hand, the distribution has very different features from those
described above. These features are as follows.
\begin{enumerate}
\item
First, the distribution is not smooth, but divided into two groups of 
heavily and lightly blended events.  The two groups are roughly divided 
by $f_{\rm B}\sim 0.5$.  This kind of distribution occurs when the number 
of blended stars in the resolution disk is of an order of one or less.  
If the number of blended stars is substantially larger (smaller) than this, 
most events will be distributed in the region of high (low) blended light 
fraction fraction.  We find that the average number of blended stars is 
$\sim 0.42$.  
\item
Another noticeable feature is that both of the events caused by stellar 
and remnant lenses have similar distributions.  This is because there is 
no contribution of the flux from the lens to the total blended flux.
\end{enumerate}

Considering the characteristics of the distributions for the stellar 
and lens blending dominated regimes, we find that the distribution of
{\it MPF} events has features of both types of blending.  For example, 
although majority of remnant events are located in the very small 
$f_{\rm B}$ region, a considerable fraction of these events are smoothly 
distributed in a wide range of blended light fraction.  In addition, 
although still events can be divided into two groups of heavily and 
lightly blended events, the distinction between the two groups is less 
clear.  These characteristics imply that both types of blending are 
important for events to be detected from the {\it MPF} survey.  This can 
be seen also in Figure~\ref{fig:three}, where we plot the distribution 
of blended light fraction.  From the figure, one finds that both types 
of blending are nearly equally important.

\section{Additional Astrometric Information}

When a source star blended with background stars is gravitationally 
magnified, the apparent position of the blended image centroid moves 
toward the source position during the event.  The amount of the 
centroid shift is 
\begin{equation}
\Delta r = \left\vert \bar{\rvec} - \bar{\rvec}_0  \right\vert
={\cal D} |\rvec_{\rm S}-\bar{\rvec}_0|;\qquad
{\cal D} = {(1-f_{\rm B})(A-1)\over (1-f_{\rm B})(A-1)+1},
\label{eq6}
\end{equation}
where $\rvec_{\rm S}$ is the position of the source, and $\bar{\rvec}_0$
and $\bar{\rvec}$ are the positions of the image centroids before and in 
the middle of lensing magnification, respectively.

The centroid shift can be measured in other way.  \citet{han00} and 
\citet{gould02} pointed out that the position of the lensed source star, 
$\rvec_{\rm S}$, instead of the centroid position of the combined image 
of the source and blend, $\bar{\rvec}$, can be directly measured on an 
image obtained by using difference imaging.  Difference imaging is an 
image-subtraction technique to provide accurate photometry and astrometry 
of variable stars in crowded fields \citep{tomaney96, alard98}.  In this 
technique, one first forms a high-quality ``template'' image.  For each 
of other images (``current'' image), one convolves the template image to 
the same seeing as the current image, translates it so that the two images 
are geometrically aligned, and linearly rescales its flux so that the two 
images are photometrically aligned as well.  When the convolved template 
image is subtracted from the current image, all that remain are a set of 
point spread functions (PSFs) at the locations of stars that experienced 
light variation.  Then, by transforming the coordinates of the source 
position measured on the subtracted image to the template image, one can 
measure the positional difference between $\rvec_{\rm S}$ and $\bar{\rvec}_0$.
The amount of the centroid shift measured in this way is related to the shift 
measured directly on two regular images by
\begin{equation}
\Delta r_{\rm DI} = |\rvec_{\rm S}-\bar{\rvec}_0|.
\label{eq7}
\end{equation}
We note that the centroid shift measured in this way is always larger
than the shift in equation~(\ref{eq6}) because ${\cal D}<1.0$.

\begin{figure}[t]
\epsscale{1.2}
\plotone{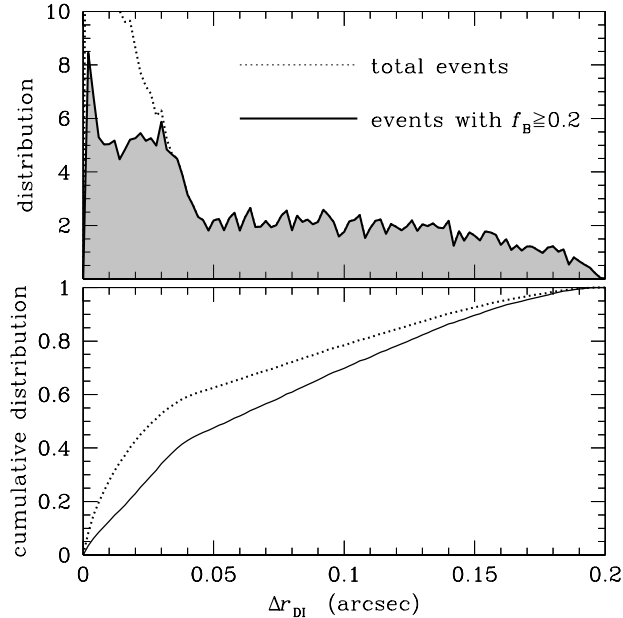}
\caption{\label{fig:four}
Distribution of source star image centroid shifts, $\Delta_{\rm DI}$ for 
microlensing events to be detectable from the {\it MPF} survey.  The dotted 
and solid curves are for all events and those with blended light fraction 
$f_{\rm B} \geq 0.2$, respectively.  Lower panel shows the cumulative 
distributions.
}\end{figure}

Astrometric information of the centroid shift might be important in 
distinguishing events blended by the lens and background stars.  This 
is because typical separation between the source and blended background 
star, which is of an order of $10^{-2}$ arcsec, is much larger than the 
separation between the source and lens, which is of an order of $10^{-4}$ 
arcsec.  Then, if no centroid shift is detected within the astrometric 
precision of the centroid shift measurement despite that blending is 
noticed from the lightcurve, the chance that the lens is the blending 
source is very high.  Figure~\ref{fig:four} shows the distribution of the 
centroid shifts for events detectable by the {\it MPF} survey.

\begin{figure}[t]
\epsscale{1.2}
\plotone{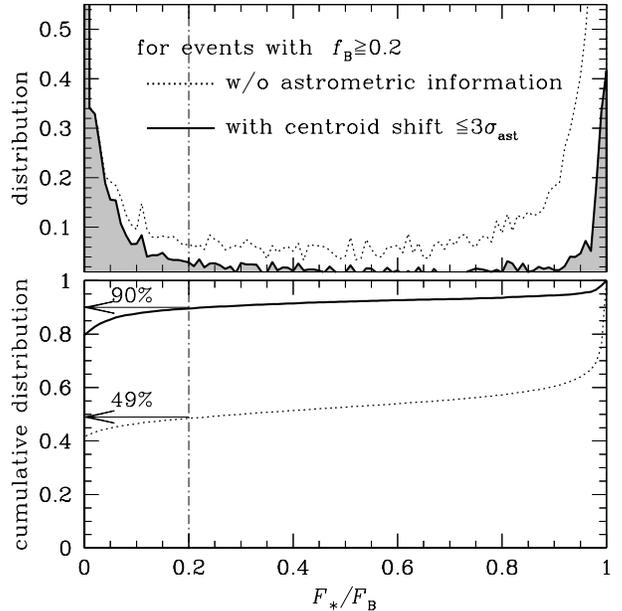}
\caption{\label{fig:five}
Distribution of the fractions of the blended flux from background star, 
$F_\star$, out of the total blended flux, $F_{\rm B}$, for {\it MPF} 
microlensing events with the blended light fraction $f_{\rm B}\geq 0.2$ 
(dotted curve).  Solid curve is the distribution for events that are 
additionally filtered out with astrometric information under the condition 
that the measured source star image centroid shift, $\Delta r_{\rm DI}$, 
is less than 3 times of the astrometric uncertainty, $\sigma_{\rm ast}$.  
Lower panel shows the cumulative distributions.
}\end{figure}

We investigate the usefulness of the astrometric information in sorting 
out events where most of the blended flux is attributable to the lens.
For this, we calculate the fraction of blended flux from background stars,  
$F_\star$, out of the total blended flux for events where blending is 
noticed with $f_{\rm B}\geq 0.2$ but no centroid shift is detected within 
the astrometric uncertainty.  The astrometric uncertainty of the centroid 
shift measurement is 
\begin{equation}
\sigma_{\rm ast} \sim {{\rm FWHM}\over N_\nu^{1/2}},
\label{eq8}
\end{equation}
where $N_\nu$ is the photon count.  We assume that the centroid shift is 
measured from difference imaging, i.e.\ $\Delta r_{\rm I} = |\rvec_{\rm S}
- \bar{\rvec}_0|$.  Despite the absence of atmosphere, difference imaging 
on space-based observation may not be simple due to variations in the 
PSF profile from image to image caused by temporal variations in the 
thermal load on spacecraft \citep{gilliland00}.  Furthermore, the {\it MPF} 
mission may suffer from undersampling of the PSF which further complicates 
the geometric alignment procedure which involve interpolation of 
undersampled PSFs.  To account for these possible additional sources of 
uncertainty in astrometric measurement, we consider only centroid shifts 
with $\Delta r_{\rm DI} > 3 \sigma_{\rm ast}$ can be firmly detected.

In Figure~\ref{fig:five}, we present the resulting distributions of the 
fraction $F_\star/F_{\rm B}$.  Dotted curve is for events with blended 
light fractions $f_{\rm B} \geq 0.2$ and solid curve is the distribution 
for events that are additionally filtered out with astrometric information 
under the condition that the measured centroid shift is $\Delta r_{\rm DI}
\leq 3 \sigma_{\rm ast}$.  We note that the blended flux from background 
stars works as a contaminant to the flux from the lens, and thus small 
ratio of $F_\star/F_{\rm B}$ implies that the blended flux is mostly from 
the lens.  From the distributions, we find that blended flux of nearly 
half (49\%) of all events with $f_{\rm B}\geq 0.2$ is contaminated by the 
flux from background stars by more than $F_\star/F_{\rm B}=20\%$. However, 
for the sample of events filtered out by using the additional astrometric 
information, it is found that only a minor fraction ($\sim 10\%$) of events 
are contaminated by background stellar blending.  We also find that this 
sample of events comprises $\sim 22\%$ of all {\it MPF} lensing events.  
Considering the total event rate of $\sim 3500$ events/yr of the survey, 
the rate of these events is $\geq 700$ events/year, which is large enough 
for the statistical analysis of the lens populations.

\section{Conclusion}

We investigated the effect of blending in future gravitational 
microlensing surveys using a 1 m class space telescope.  For this, we 
carried out simulation of Galactic bulge microlensing events to be 
detected from the proposed {\it MPF} lensing survey by imposing 
realistic observational conditions and detection criteria.  From this 
simulation, we found that the contribution of the flux from blended 
background stars to the total blended flux is equivalent to that of the 
lens itself, implying that characterizing lenses from the analysis of 
the blended flux would not be easy.  However, with the additional 
astrometric information of the source star image centroid shift, we found 
that it would be possible to isolate events for which most of the blended 
flux is attributable to the lens.  For the sample of events obtained by 
imposing a criterion that the centroid shift should be less than three 
times of the astrometric uncertainty among the events for which blending 
was noticed with blended light fractions $f_{\rm B}>0.2$, we estimated 
that the contamination of the blended flux by background stars would be 
less than 20\% for most ($\sim 90\%$) of the sample events.  The expected 
rate of these events is $\gtrsim 700$ events/yr, which is large enough for 
the statistical analysis of the lens populations.

\acknowledgments
This work was supported by the grant (C00072) of the Korea 
Research Foundation.

\end{document}